\documentclass[twocolumn,runningheads,natbib]{svjour2}
\smartqed  % flush right qed marks, e.g. at end of proof

\usepackage{graphicx}

\journalname{Astrophysics and Space Science}

\begin{document}

\title{Search for fast optical activity of SGR 1806-20 at the SAO RAS 6-m 
telescope\thanks{This work has been supported by the Russian Foundation 
for Basic Research (grant No 04-02-17555), Russian Academy of Sciences 
(program "Evolution of Stars and Galaxies"), and by the Russian Science 
Support Foundation. Also the authors are thankful to the anonymous referee for 
his/her valuable comments.}}

\titlerunning{Search for optical activity of SGR 1806-20}        % if too long for running head

\author{G. Beskin \and V. Debur \and V. Plokhotnichenko \and S. Karpov
\and A. Biryukov \and L. Chmyreva \and A. Pozanenko \and K. Hurley
}
\authorrunning{G.Beskin et al} % if too long for running head

\institute{G. Beskin \at
              Special Astrophysical Observatory of RAS \\
              Nizhniy Arkhyz, Karachaevo-Cherkessia, Russia, 369167 \\
              \email{beskin@sao.ru} 
}

\date{Received: date / Accepted: date}

\maketitle

\begin{abstract}

The region of SGR 1806-20 localization was observed during its
gamma-ray activity in 2001. The observations have been performed on the
6-meter telescope of the Special Astrophysical Observatory,
using the Panoramic Photometer-Polarimeter (PPP). The search for
variability was performed on the $10^{-6}$ - $10$ s time scale, and its 
results were compared to the properties of corresponding x-ray flares.

\end{abstract}

\section{Introduction}

There is now increasing evidence that the soft repeater SGR 1806-20 hosts
a magnetar, i.e. a neutron star with an anomalously high magnetic field 
$B>10^{14}$ Gs \citep{duncan_1992,woods_2004}. The SGR 1806-20 is the most 
active among soft gamma repeaters and is characterized by the emission of 
short sporadic flashes of soft gamma rays with characteristic durations of 
10 ms - 1 s and luminosities of $10^{39}$ - $10^{42}$ erg/s. 
\citep{gogus_2001, hurley_2000}.
They are detected during sporadic periods of the object's activity lasting 
from days to months. Pulsations of the persistent X-ray flux of SGR 1806-20
with a period of 7.47 s were discovered \citep{kouv1998}.
The culmination of the long period of its activity that started in the end
of 2003 was a giant flare of 27th December 2004 
\citep{bork_2004, maz_2004, golen_2004}. During the main spike that lasted 
0.2-0.5 s, about $10^{47}$ ergs above 50 Kev was emitted for a distance of 
15 kpc \citep{palmer_2005, mereg_2005, shwartz_2005, teras_2005}.
The long part of the flare showed a pulsation with the period of 7.57 s 
during about 300 s \citep{bork_2004}.

Unfortunately, even the minimal estimation of the distance to SGR 1806-20 is 
6 kpc (while the most reasonable one is 15 kpc), and the absorption reaches 
$A_V\sim30$ $m$ \citep{eikenberry_2004,mcclure_2005}, so there is not 
much hope of detecting its optical emission, even though its infrared 
counterpart seems to have been found, with a magnitude of $K=21.6$ $m$ 
\citep{israel_2005}.

However, we have carried out a set of observations of the location of SGR 
1806-20 in the optical band with 1 $\mu$s temporal resolution to try to 
detect very short and strong optical spikes. The epoch of the observations 
has been chosen in relation to the increase of $\gamma$-ray activity of the 
source, according to HETE data \citep{gcn_1068,gcn_1089}.

We have reported the preliminary results of our monitoring in earlier papers 
\citep{beskin_gcn,beskin_2003}. In the present work we describe the process 
of observation, the equipment, and we present a more detailed analysis of 
the acquired data. 

\section{Observations, hardware and software}

The field of SRG 1806-20 has been observed using the 6-meter telescope of the 
Special Astrophysical Observatory. Observations were carried out with 
Panoramic Photometer-Polarimeter (PPP) with high time resolution 
\citep{plokhotnichenko} in the telescope prime focus on June 20 2001 
(2 days after HETE trigger), and on August 22 2001 (15 hours after HETE 
trigger) \citep{gcn_1068,gcn_1089}.

The main part of PPP is the Positional Sensitive Detector (PSD), which 
consists of a vacuum tube with a standard S20 photocathode, a set of 
microchannel plates and a four-electrode anode. The pixel size of the 
detector is $0''.21$, the FOV is a circle of about $1'.5$, and the time 
resolution (dead time) is 1 $\mu$s \citep{debur}.

For the determination of the photon arrival times and coordinates, and for 
the storage of the whole data set, a special ``time-code'' converter 
``Quantochron 4-48'' connected to the PC in realtime has been used.
The information on the observations is summarized in Table 
\ref{table_observations}.

\begin{figure}
\centering
{\centering \resizebox*{1\columnwidth}{!}{\includegraphics[angle=0]{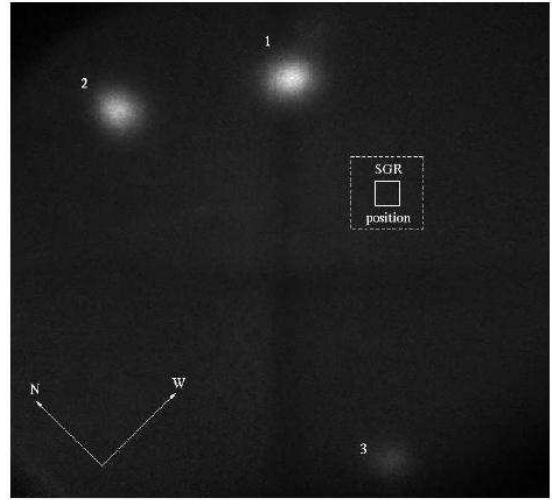} }\par}
\caption{Observed field of SGR 1806-20. Numbered stars are listed in Table \ref{table_stars}.}
\label{fig_field}
\end{figure}

The search for variability has been performed in the region centered on the 
IPN localization of the source \citep{hurley_1999}. The statistical 
properties of the photon lists have been studied separately for 9 square 
boxes with the size of $6''.5$ covering the localization region 
(see Fig.\ref{fig_field}).

As an indicator of the variability we used the function $y_2$, defined as 
the ratio between the distributions of the photon arrival time intervals for 
the source and background \citep{shvarts_1977}. It is:
\begin{equation}
y_2(\tau)=\frac{P_s(\tau)-P_b(\tau)}{P_b(\tau)},
\label{eq_y2}
\end{equation}
where $\tau$ is the interval between the times of arrival of successive photons, 
while $P_s(\tau)$ and $P_b(\tau)$ are the distribution functions of 
$\tau$ for the source and background boxes, respectively.

We used the following definitions and relations: $I(t)$ - intensity variation 
during the flare and its mean value:
\begin{equation}
<I> = \frac{1}{<\tau>} = \frac{1}{T}\left[I_0 + \int_0^T{I(t)dt}\right], 
\end{equation}
where $\tau$ is the mean interval between the arrival times of
successive photons,  $T$ - mean time between the flares and 
$I_0$ - persistent emission intensity; $I_{max}$ - maximum intensity of the 
flare; characteristic time scale of the variability:
\begin{equation}
\tau_{var} = \frac{1}{I_{max}}\int_0^T{I(t)dt}; 
\end{equation}
relative amplitude of the intensity variations $A \equiv \frac{I_{max}}{<I>}$, 
and relative power of the variable emission component
\begin{equation}
S = \frac{1}{<I>T}\int_0^T{I(t)dt} = \frac{I_{max}\tau_{var}}{<I>T} = FA,
\end{equation}
where $F = \tau_{var}/T$ - the flares duty cycle.

The function $y_2$ is related to the variability parameters defined
above \citep{shvarts_1977} as
\begin{equation}
y_2(\tau \ll <\tau>, \tau_{var}) \approx \frac{kS^2}{F}
\label{eq_y2_2}
\end{equation}
where $k = \frac{1}{<I>T}\int_0^T{I^2(t)dt}$ - flare shape cofficient.
For example, sinusoidal flares have $k = 0.25$, triangular ones have
$k = \frac{2}{3} - 2\beta + 2\beta^2 \approx 0.2 \div 0.7$.

From the registered photon lists it is easy to compute $y_2(\tau)$, from 
which $S$ follows:
\begin{equation}
S(\tau) \approx \sqrt{\frac{y_2(\tau) F}{k}}
\label{eq_S}
\end{equation}
The absence of variability is rejected if $S(\tau)>3\sigma_S(\tau)$, where
\begin{equation}
\sigma_S(\tau) = \left(\frac{1}{P_b(\tau)N}\right)^{\frac14},
\label{eq_sigma}
\end{equation}
and $N$ is the number of photons detected 
\citep{plokhotnichenko_disser, plokhotnichenko_comm}.

\section{Results of observations}

Using the arrival times of $10^6$ photons, detected in each sub-box of the 
SGR 1806-20 localization, the distribution function of intervals between 
the following photons has been built. The mean background intensity has 
been $152$ counts/s with nearly Poissonian statistics.

\begin{figure*}
\centering
{\centering \resizebox*{2\columnwidth}{!}{\includegraphics[angle=270]{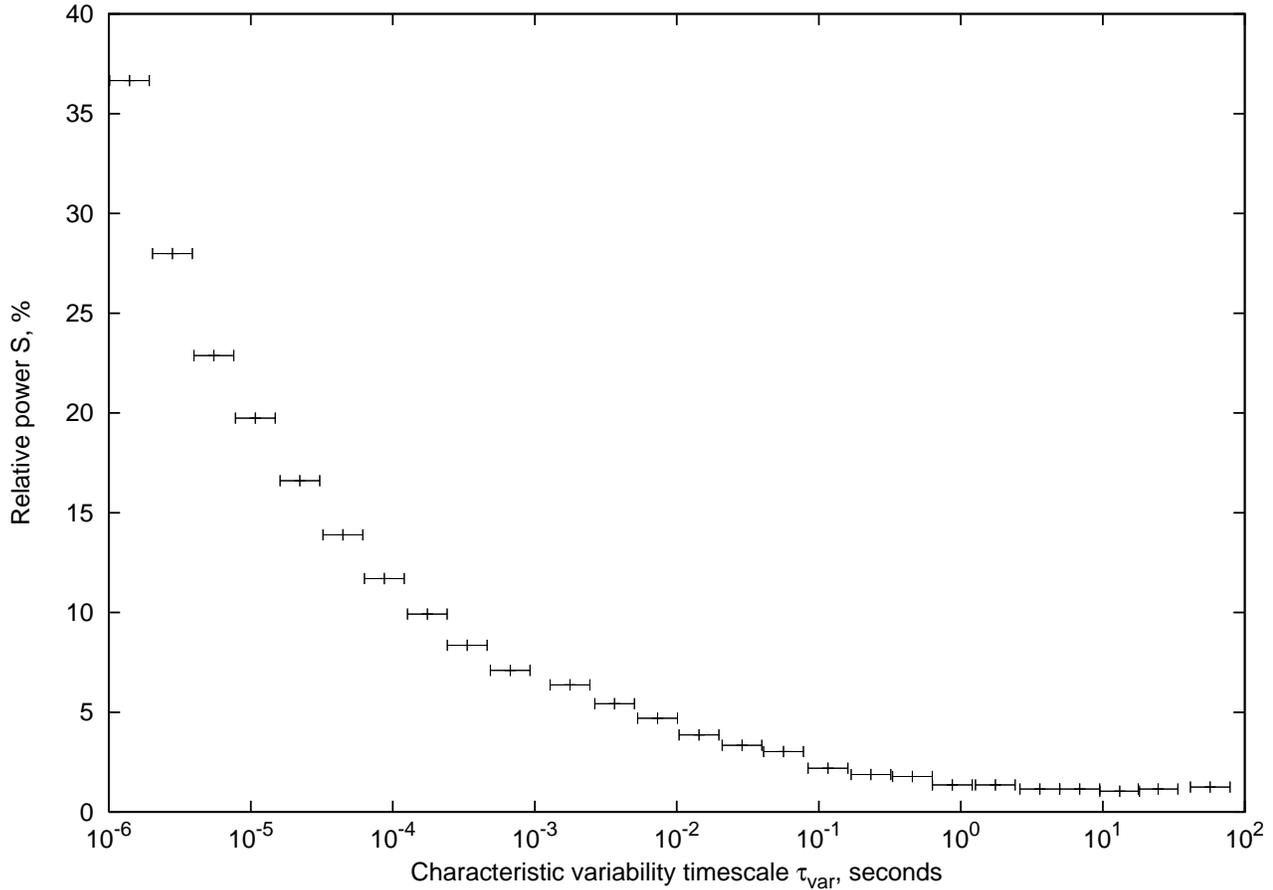}} \par}
\caption{Upper limits for the relative power of the variable emission.}
\label{fig_limits}
\end{figure*}

Fig.\ref{fig_limits} shows the upper limits for the relative power $S$ of 
the variable emission component derived using equations (\ref{eq_y2}) - 
(\ref{eq_sigma}). The model of triangular flares with the duty cycle of 
$0.1$ was used. 

We used the information on flares detected in gamma and X-ray bands (see, 
for example, \cite{woods_2004} and \cite{gotz_2006}).

For the photometric calibration, the stars listed in Table \ref{table_stars}
have been used.

\begin{table}
\caption{Observations details}
\begin{tabular}{ccccc}
\hline
\noalign{\smallskip}
Date  &Start Time  &Exposure        &Zenith & Filter\\
      &(UT)        &Time (s)        &Distance       &\\
\noalign{\smallskip}
\hline
\noalign{\smallskip}
20.06.2001 &$19^{h}58^{\rm m}$ & 4544.6&$67.^{\circ} 1$&B\\
22.08.2001 &$20^{h}31^{\rm m}$ &800.33&$77.^{\circ} 7$&B\\
\noalign{\smallskip}
\hline
\end{tabular}
\label{table_observations}
\end{table}

\begin{table}
\caption{USNO-A2.0 stars in the field of SGR 1806-20}
\begin{tabular}{ccccc}
\hline
\noalign{\smallskip}
N   &R.A.           &DEC.   &Bmag   &Rmag   \\
&J2000*          &J2000  & &\\
\noalign{\smallskip}
\hline
\noalign{\smallskip}
1    &$18^{h}08^{\rm m}39.^{\rm s} 42$&$-20^{\circ}24^{\prime} 11.^{\prime\prime}74$&14.6  &13.7\\
2    &$18^{h}08^{\rm m}41.^{\rm s} 13$&$-20^{\circ}24^{\prime} 02.^{\prime\prime}29$&15.5  &14.6\\
3    &$18^{h}08^{\rm m}41.^{\rm s} 47$&$-20^{\circ}25^{\prime} 12.^{\prime\prime}61$&16.5  &15.9\\
\noalign{\smallskip}
\hline
\end{tabular}
*Monet D.G., 1998, BAAS 30, Vol.4, 120.03
\label{table_stars}
\end{table}

So, the upper limits for flashes with the duration of $\sim 10^{-2}$ s 
(similar to the parameters of $\gamma$-ray flashes) will be about 
$20$ $m$. For very rare flashes with the duty cycle of $10^{-4}$, 
the upper limits will be thirty times better - we should be able to detect
events of $23$ $m$ $\div$ $24$ $m$.

Unfortunately, even having good upper limits for the variable optical 
component, due to huge absorption it is impossible to derive a reasonable 
estimate of the luminosity. Assuming $A_V\sim20$ $m$ and a distance of 6 
to 15 kpc \citep{eikenberry_2004,mcclure_2005}, we have for the upper limits 
of flaring optical luminosity of SGR 1806-20 values ranging from 
$7\cdot10^{46}$ erg/s to $4\cdot10^{47}$ erg/s. These values are similar to
the peak luminosity of the object in the $\gamma$-ray band, i.e. to the 
values registered during the giant flare. This means that very short and 
very rare ``giant optical flares'' could be detected.

\end{document}